\documentclass[10pt, two column, twoside]{IEEEtran}
\usepackage{color}
\usepackage[colorlinks, linkcolor=color1, anchorcolor=blue, citecolor=color1]{hyperref}

\usepackage{hyperref}
\usepackage{cite}

\usepackage{enumitem}

\usepackage{amssymb}
\usepackage{amsmath}
\usepackage[lined,boxed,commentsnumbered, ruled]{algorithm2e}
\usepackage{mathrsfs}
\usepackage{algorithmic}
\usepackage{bm}

\usepackage{pgfplots}
\usepackage{tikz}
\usetikzlibrary{arrows}
\usepackage{subfigure}
\usepackage{graphicx,booktabs,multirow}

\definecolor{colorhkust}{RGB}{20,43,140}
\definecolor{colortsinghua}{RGB}{116,52,129}
\definecolor{color1}{RGB}{128,0,0}

\usepackage[normalem]{ulem}

\renewenvironment{IEEEbiography}[1]
  {\IEEEbiographynophoto{#1}}
  {\endIEEEbiographynophoto}

\date{}

\begin{document}

\title{Federated Machine Learning for Intelligent IoT via Reconfigurable Intelligent Surface}
\author{Kai Yang,  Yuanming~Shi, Yong Zhou, Zhanpeng Yang, Liqun Fu, and Wei Chen 
        \thanks{K. Yang, Y. Shi, Y. Zhou and Z. Yang are with
ShanghaiTech University; L. Fu is with Xiamen University; Wei Chen is with Tsinghua University. }
                }
                           
\IEEEpeerreviewmaketitle


\maketitle

\begin{abstract}
Intelligent Internet-of-Things (IoT) will be transformative with the advancement of artificial intelligence and high-dimensional data analysis, shifting from ``connected things" to ``connected intelligence". 
This shall unleash the full potential of intelligent IoT in a plethora of exciting applications, such as self-driving cars, unmanned aerial vehicles, healthcare, robotics, and supply chain finance. 
These applications drive the need of developing revolutionary computation, communication and artificial intelligence technologies that can make low-latency decisions with massive real-time data. 
To this end, federated machine learning, as a disruptive technology, is emerged to distill intelligence from the data at network edge, while guaranteeing device privacy and data security. 
However, the limited communication bandwidth is a key bottleneck of model aggregation for federated machine learning over radio channels. 
In this article, we shall develop an over-the-air computation based communication-efficient federated machine learning framework for intelligent IoT networks via exploiting the waveform superposition property of a multi-access channel. 
Reconfigurable intelligent surface is further leveraged to reduce the model aggregation error via enhancing the signal strength by reconfiguring the wireless propagation environments. 
\end{abstract}


\section{Introduction}
Internet of Things (IoT) is envisioned to enable automated data transmission and offer ubiquitous wireless connectivity for trillions of devices (e.g., smart phones and sensors) with the capabilities of sensing, communication, computation and control. The vast amount of data generated by IoT devices can be exploited to extract useful information by machine learning, thereby enabling various intelligent IoT services. The emerging intelligent IoT applications include self-driving cars, unmanned aerial vehicles (UAVs), robotics, healthcare, and supply chain finance, etc. However, enabling the paradigm shift from ``connected things" to ``connected intelligence" in the era of 6G via modern machine learning techniques \cite{letaief2019roadmap} faces three main challenges among others. First, transmitting data with private information to the cloud server is susceptible to eavesdropping and data modification attacks. Second, aggregating a large volume of distributed data for model learning over radio channels is likely to cause network congestion and lead to excessive network latency, due to limited spectrum resources. Third, many connected devices have strong computation capabilities, which have not been fully exploited to collaboratively train sophisticated models with high performance and accuracy requirements \cite{Dong_IEEEnet18}.

Federated machine learning \cite{yang2019federated} is a promising solution for privacy-sensitive and low-latency intelligent IoT applications and has the capability of harnessing distributed computation resources \cite{Xu_edgeai19,kang2020incentive}. 
Specifically, federated machine learning allows each IoT device to keep its data locally and only requires each IoT device to upload its locally updated model to the edge aggregation server during model training via wireless links. 
This prevents revealing the collected data at the IoT devices to other devices and the aggregation server, thereby enhancing device privacy and data security. 
The shared global model with high prediction accuracy can be learned at the edge aggregation server by coordinating a fleet of IoT devices to participate model aggregation.
The learning latency can be significantly reduced by leveraging advanced federated machine learning techniques to avoid offloading data to the cloud data center.
Although more selected devices can yield better model aggregation performance, periodical model updating from the selected IoT  devices becomes the key communication bottleneck to unleash the full potential of federated machine learning for intelligent IoT applications. 

Over-the-air computation (AirComp) provides a novel simultaneous access technique to support fast local model aggregation for federated machine learning via exploiting the signal superposition property of multi-access channels \cite{yang2018federated,zhu2018low,amiri2019federated}. This can be achieved by concurrent transmissions of analog locally updated models at each device, followed by directly receiving the average of local models at the edge server. This perfectly matches the functional computation of a weighted average of the local model updates for model aggregation to update a global model at the edge server in federated machine learning. Instead of treating the signals from other devices as interference in conventional multiple access schemes for mobile data services, AirComp is able to harness interference to reduce the spectrum resource requirements in the learning process over radio channels. However, the performance of AirComp is still limited by the unfavorable wireless propagation conditions.  

\begin{figure*}[ht]
  \centering
  \includegraphics[width=0.8\linewidth]{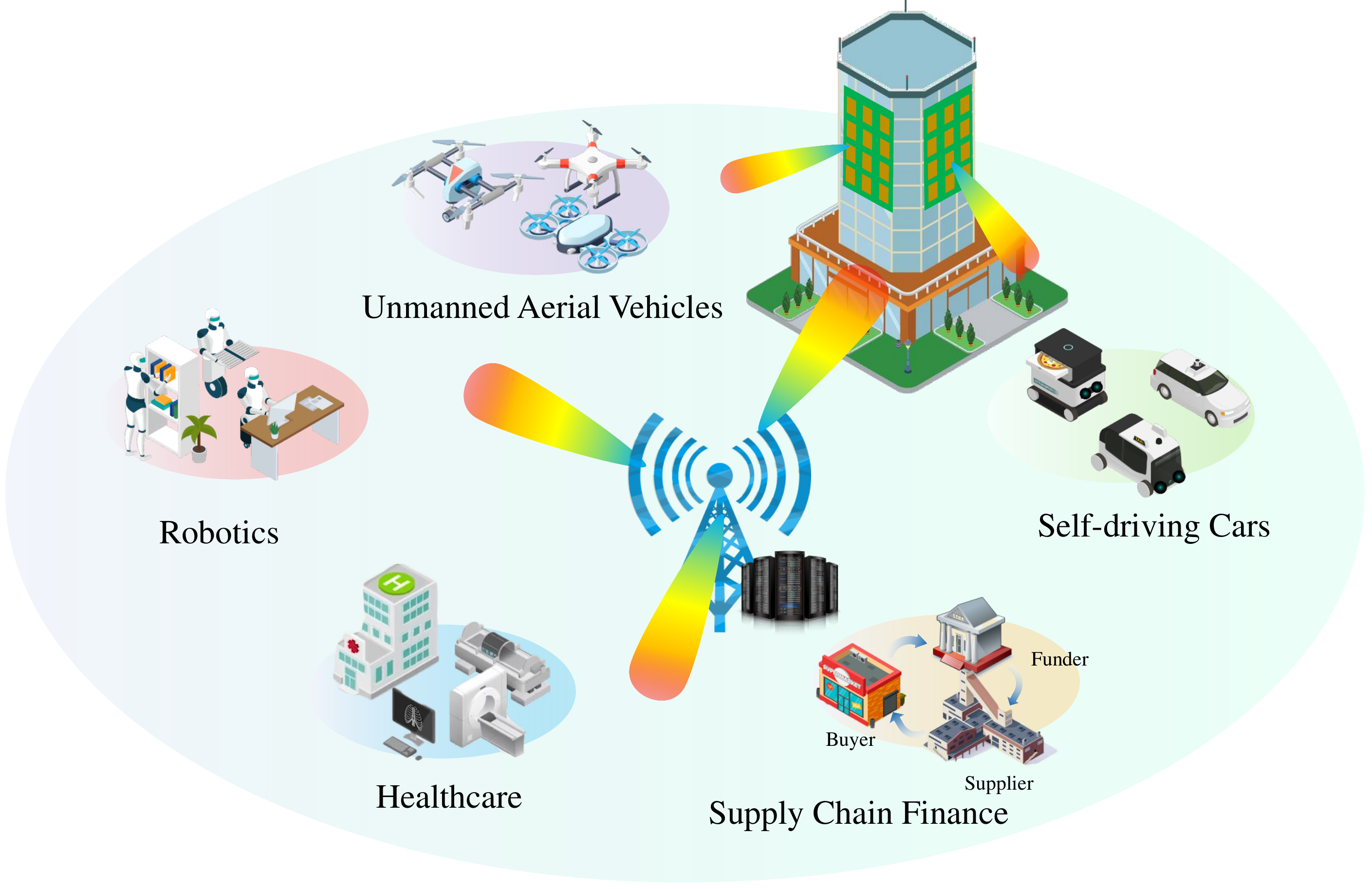}
  \caption{Federated machine learning for intelligent IoT applications via reconfigurable intelligent surface.}
  \label{fig_systemArchitecture}
\end{figure*}

Reconfigurable intelligent surface (RIS) \cite{Rui_ComMag19}, emerging as a cost-effective technology, has the potential to further reduce the model aggregation error of AirComp by reconfiguring the wireless propagation environments. 
An RIS is typically a flat artificial metasurface consists of many passive reflecting elements with adjustable phase shifts, each of which is software-controlled by a smart controller. Through jointly controlling all reflecting elements, RIS is able to introduce a desired phase shift on the incident signals, which can be exploited to enhance the signal power and mitigate the co-channel interference. 
In addition, it turns out that RIS has the huge potential to be integrated with many emerging technologies, including edge machine learning, multiple-input multiple-output communications \cite{Huang_TWCRIS19}, Terahertz communication, and sparse-code multiple access \cite{han2018optimal}. 
In this article, we shall develop a novel simultaneous access scheme empowered by RIS to boost the performance of  model aggregation, thereby  designing a communication-efficient federated machine learning framework for intelligent IoT.

The remainder of this article is organized as follows. We provide an overview of federated machine learning for intelligent IoT applications in Section II. In Section III, we present the design of AirComp for fast model aggregation of federated machine learning. Section IV discusses RIS-empowered model aggregation for achieving communication-efficient federated machine learning. Finally, Section V concludes this article. 

\section{Federated Machine Learning Meets Intelligent IoT}
In this section, we shall describe the principles, present the potentials, and discuss the challenges of federated machine learning for intelligent IoT. 

\subsection{Federated Machine Learning for Intelligent IoT Services} 
The machine learning technologies become indispensable for enabling intelligent IoT services by distilling intelligence from the data, such as self-driving cars, UAVs, healthcare, etc.
In the conventional cloud-based machine learning systems, the data generated or collected by a large number of devices need to be transmitted to a centralized cloud server, which is competent to train an artificial intelligence (AI) model. 
With the growth of computation power of IoT devices and due to the increasing concern over data security and privacy, pushing the AI engine from the cloud server to the network edge has attracted enormous academic and industrial efforts. Federated machine learning provides a novel solution for collaboratively training a global machine learning model by leveraging the distributed computation resources across IoT devices, while always keeping the data locally at devices.

A federated machine learning system for intelligent IoT applications normally consists of a central aggregation server (e.g., base station and access point) and a fleet of IoT devices, as shown in Fig. \ref{fig_systemArchitecture}. 
The aggregation server maintains and updates a global model, while each IoT device trains a local model. 
The global and local models are iteratively updated through a number of communication rounds until the consensus of the global model is reached. 
As illustrated in Fig. \ref{fig_FLsystem}, the following three steps are sequentially performed during each learning round:
\begin{figure*}[h]
  \centering
  \includegraphics[width=0.8\linewidth]{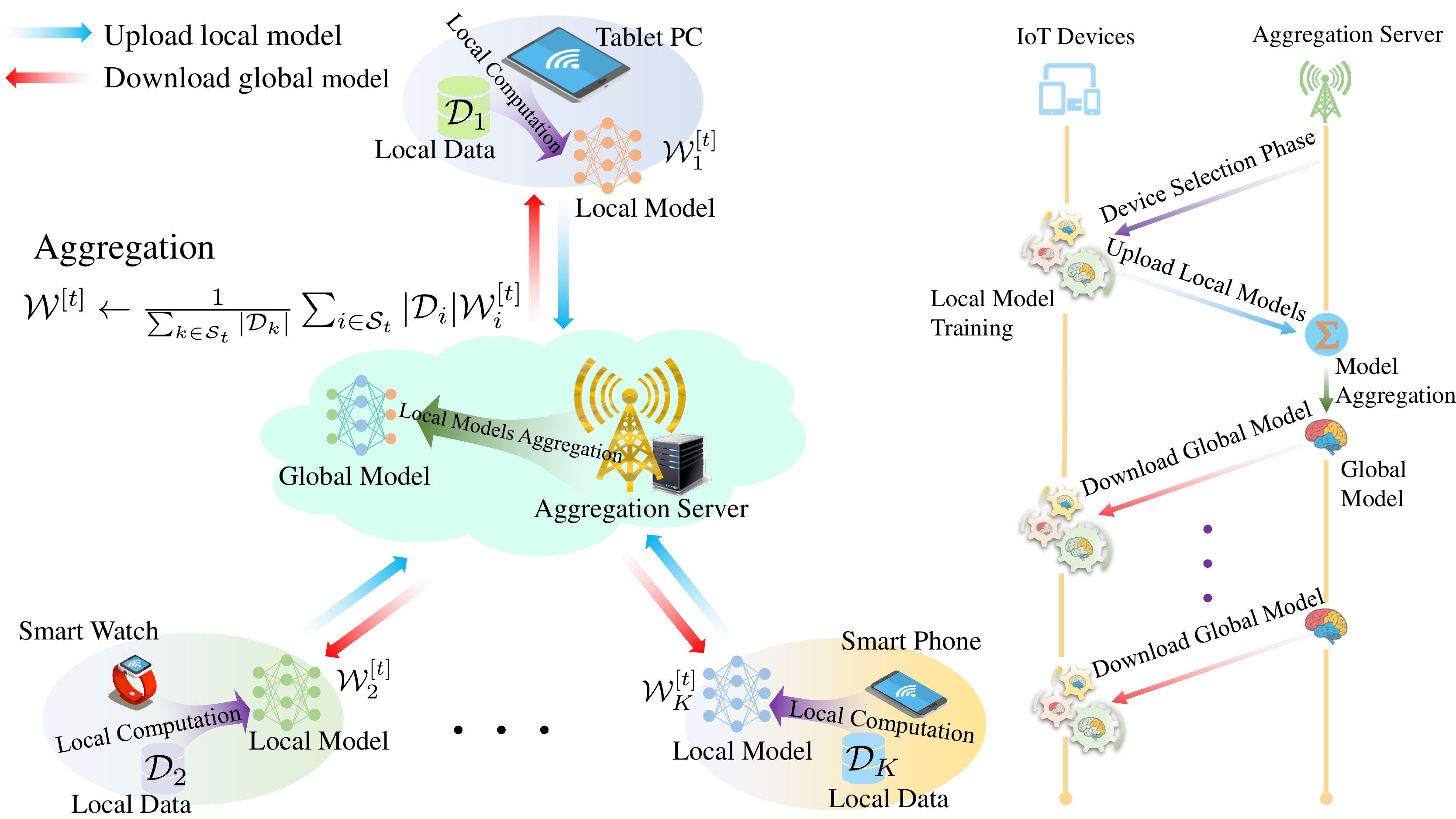}
  \caption{Typical system architecture and communication process of federated machine learning in intelligent IoT.}
  \label{fig_FLsystem}
\end{figure*}

$\bullet$ Device selection: to reduce the communication overhead, the aggregation server only selects a subset of IoT devices (e.g., with good channel conditions and important model updates) for local model training and updating. 

$\bullet$ Local model upload:  based on the most recently downloaded global model, each IoT device updates its local model according to the learning algorithms (e.g., stochastic gradient method), followed by transmitting the local updates to the aggregation server via the wireless uplink transmission for global model updating.

$\bullet$ Global model download: the server aggregates the local models from the selected IoT devices to update the  global model, which is then broadcasted to the selected IoT devices via downlink transmissions for next round learning. 

It is worth noting that the data generated or collected at a single IoT device are very limited to train a high performance AI model. 
To address the challenge, federated machine learning is capable of overcoming the data isolated islands problem in intelligent IoT systems, where a vast amount of privacy-sensitive data located at different IoT devices can be leveraged to train a common global AI model. 
The federated machine learning in intelligent IoT networks has the following advantages: 

$\bullet$ Preserving device privacy and data security: 
without sharing the collected data between the central server and the IoT devices, the raw data of each IoT device are not exposed to other devices as well as the central server. 
Hence, federated machine learning guarantees the device privacy and data confidentiality for intelligent IoT applications. 

$\bullet$ Enabling collaborative training: federated machine learning is capable of pooling the computation resources and the collected data over a large number of distributed IoT devices for collaboratively building a  global machine learning model. 
As a result, each device obtains a better machine learning model than that can be achieved by itself. 

$\bullet$ Reducing network latency: without requiring the IoT devices to upload the huge amount of training data to the remote cloud center, federated machine learning avoids the radio transmission of potentially heavy communication loads and only needs to communicate with the proximal edge server, thereby significantly reducing the network latency. 

These advantages make federated machine learning a promising solution that exploits computation and communication resources for distilling intelligence directly at network edge, thereby unleashing the full potential of machine learning in a plethora of exciting intelligent IoT applications.

\subsection{Key Intelligent IoT Applications Empowered by Federated Machine Learning}
Federated machine learning is a promising technology that is expected to play a pivotal role in various intelligent IoT applications to provide low-latency decisions with strong privacy and security guarantees, including self-driving cars, UAVs, healthcare, robotics, and supply chain finance, as shown in Fig. \ref{fig_systemArchitecture}. We shall introduce some typical applications in different industries as follows:

\subsubsection{Self-Driving Cars}
IoT is an indispensable driving force behind the fast growing self-driving car industry. 
Embedded with hundreds of sensors (e.g., video camera, lidar, and radar), each future self-driving car is predicted to generate 100 gigabytes data per second. 
Those data can be exploited by cutting-edge machine learning technologies to enable many crucial functions, including adaptive obstacle avoidance and pace adaptation according to the environments. 
The self-driving cars basically require quick response to complicated situations, which pose stringent latency requirements for performing intelligent tasks \cite{Jun_IEEEPro19}. 
In addition, data security is critical for self-driving cars since the data of vehicles usually contain lots of sensitive information of the users. Federated machine learning presents promises in efficiently and securely training AI models across smart vehicles and reduces the network latency by exploiting intelligence at network edge.

\subsubsection{Unmanned Aerial Vehicles}
In recent years, UAVs have found a variety of applications in civilian and commercial domains such as traffic monitoring, cargo delivery, and virtual reality (VR). It is expected that UAVs will be provided with ubiquitous wireless connectivity and integrated intelligence in future communication systems to support more intelligent IoT applications \cite{Bennis_IEEEPro19}. 
Performing AI tasks in UAV networks may introduce large communication and signaling delay as constantly communicating with the remote cloud center is required. Federated machine learning thus shows promises in reducing the network delays by pushing the AI engine to the network edge and collaboratively training AI models across a large number of UAVs.

\subsubsection{Healthcare}
Machine learning with real-world healthcare data has shown its great potential in improving quality of healthcare, such as generating diagnostic tools and predicting disease risks. 
Healthcare data are particularly sensitive and private for patients, which are protected by the strict regulatory policies over the world \cite{yang2019federated}. To overcome the isolation issue of data, federated machine learning shows its potential in collaboratively learning AI models for intelligent healthcare services while guaranteeing the data privacy. Moreover, the local updates can be further encrypted (e.g., using homomorphic encryption) before transmission to protect against information leakage. NVIDIA Clara is a representative healthcare service platform featured by federated machine learning for protecting patient data in hospitals and medical institutions. More recently, Standford Institute of Human-Centered AI develops an in-home system powered by federated machine learning to monitor resident for coronavirus symptoms in the midst of the COVID-19 pandemic.

\subsubsection{Robotics}

The Internet of Robotic Things (IoRT), integrating robotics and IoT, empowers the intelligent robots monitor the surrounding events, make immediate decisions, and take appropriate actions. 
Being able to interactively react to unexpected events, IoRT has wide applications in many manufacturing industries, including precision agriculture and industrial IoT. 
A single robot normally makes decisions based on the local observations and limited intelligence capability, which leads to excessive decision-making delay and inaccurate reactions to the dynamic environments. 
Federated machine learning has the great potential to fully leverage the computation capabilities of distributed robots to achieve collaborative intelligence, thereby enhancing the capability of performing more complex and challenging interactive tasks.

\subsubsection{Supply Chain Finance}

Supply chain finance integrated with machine learning and IoT is capable of speeding up the capital and information flows throughout the supply chain, thereby reducing the financial gap between the buyers and suppliers. 
In particular, the data from industrial monitoring in supply chain systems need to be transferred among different business entities and can be exploited to identify and extract underlying patterns with cutting-edge machine learning technologies. 
Being highly sensitive and also the key interests of business entities, the supply chain data however are vulnerable to security issues in such a distributed  system. 
Federated machine learning provides a new solution to avoid information leakage as well as reduce the performance risk and the credit risk for supply chain finance.

\subsection{Communication Challenge of Federated Machine Learning}
Although presenting great potentials for intelligent IoT, federated machine learning still encompasses a number of challenges, such as limited battery power, unbalanced number of data, heterogeneous computational capabilities, etc. 
In particular, communication bottleneck becomes one critical challenge in federated machine learning for intelligent IoT applications due to the limited radio resources and the iterative transmission of high-dimensional model-update parameters. 
Specifically, the learning process of federated machine learning typically incurs hundreds of communication rounds, in each of which many devices update local models through a wireless multi-access channel. 
The communication loads for iterative model updates grow linearly with the number of involved IoT devices when the conventional orthogonal multiple access scheme is adopted. 
A high volume of communication loads may lead to severe network congestion, which in turn incurs excessive network latency. 
To tackle the communication bottleneck during the model aggregation period in federated machine learning over radio channels, we shall present a new simultaneous access scheme based on AirComp to reduce the required radio resources by integrating the computation and communication. The reconfigurable intelligent surface is further leveraged to boost the performance of AirComp via alleviating the unfavorable wireless propagation conditions. 

\section{Over-the-air Computation for Federated Machine Learning}
This section presents a novel multiple access scheme integrating communication and computation, i.e., AirComp, for fast model aggregation in federated machine learning. 

\subsection{Principles of Over-the-Air Computation}
\begin{figure*}[h]
  \centering
  \includegraphics[width=0.8\linewidth]{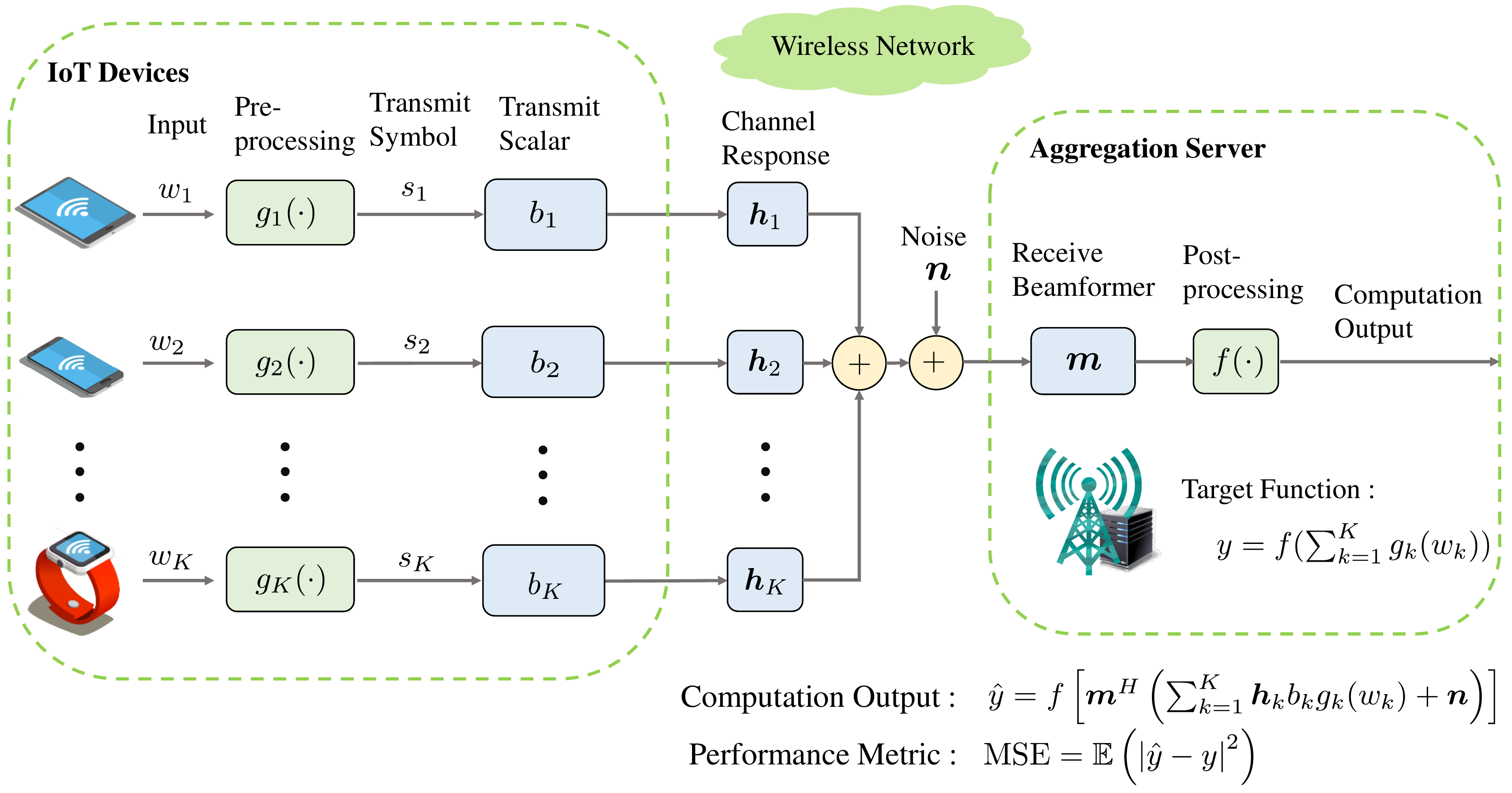}
  \caption{Illustration of over-the-air computation.}
  \label{fig_aircomp}
\end{figure*}

For computing a function value of the data at distributed mobile devices, i.e., wireless data aggregation, the conventional approach that separates communication and computation normally yields tremendous communication overheads. AirComp turns out to be an effective wireless data aggregation approach via exploiting the signal superposition property of the wireless multiple access channel. The class of target functions that can be computed over-the-air is called the nomographic function. Basically, a nomographic function can be evaluated by pre-processing the input data at each device, and then taking the sum of all data after preprocessing, followed by post-processing the summation. Fig. \ref{fig_aircomp} illustrates the procedures of computing a nomographic function from distributed IoT devices using the AirComp approach. Specifically, the input data after pre-processing is encoded as the transmit symbol at each device. By transmitting the encoded information symbols at all IoT devices over the wireless multiple access channel, the sum of all pre-processed input data can be directly decoded at the receiver with carefully designed transmit and receive strategies. The target function can be obtained by post-processing the sum of all pre-processed input data. 

The idea of joint communication and computation design approach for sensing via AirComp was first proposed for functional computation over wireless multiple access channels. It demonstrates that interference can be harnessed for computing functional values instead of being canceled, thereby significantly reducing the radio resource requirements. 
Recent years have witnessed enormous research efforts on AirComp from different aspects. Research works from the information theoretic view have studied the structured code design for different distributions of the input data, including independent Gaussian, correlated Gaussian, sum of independent Gaussian, etc. 
There have also been a number of works focusing on the transmitter and receiver beamforming design for reducing the aggregation error of AirComp. In addition, the uncertainty of the channel state information has been considered for robust AirComp. 
The synchronization issues for the implementation of AirComp have also been studied from theory to practice. 

\subsection{Over-the-Air Computation for Fast Model Aggregation}
The target of model aggregation in federated machine learning for intelligent IoT applications is to compute the weighted average of the locally computed model updates at the selected IoT devices, as illustrated in Fig. \ref{fig_FLsystem}. Fortunately, the weighted average function falls into the category of nomographic functions that can be computed based on the principles of the AirComp. Therefore, the fast global model aggregation can be achieved via harnessing the interference with much less communication bandwidth, which motivates the AirComp based fast model aggregation approach in federated machine learning. Due to the existence of noise, there are however inevitably function distortions with AirComp.

A key observation of federated machine learning is that model aggregation with large distortions could induce a notable performance drop of the model prediction accuracy. The mean-squared-error (MSE) is usually adopted as the performance metric to characterize the model aggregation error. Another key observation of federated machine learning is that more involved devices at each training round could accelerate the convergence rate of the model training process. Based on these two key observations of federated machine learning, a joint device selection and receiver beamforming design problem is formulated in \cite{yang2018federated} for fast model aggregation of federated machine learning, thereby simultaneously improving the model accuracy and model training speed. 

However, this problem turns out to be a highly intractable mixed combinatorial optimization problem with nonconvex quadratic constraints for model aggregation error reduction and the combinatorial objective function for device selection. To address this challenge, a sparse and low-rank optimization approach with novel difference-of-convex-functions (DC) programming algorithm is then developed in \cite{yang2018federated} with considerable performance improvements. Although the AirComp approach with the novel DC algorithm significantly improves the performance of model aggregation for federated machine learning, it may still suffer from unfavorable signal propagation conditions of wireless links, such as deep fading. In the next section, we shall propose to adopt the RIS to develop a smart radio environment, thereby further enhancing the performance of model aggregation in federated machine learning.

\section{Reconfigurable Intelligent Surface Empowered Over-The-Air Computation}
In this section, we propose to utilize RIS to further reduce the model aggregation error for AirComp-based federated machine learning. This is achieved by adaptively shaping the wireless propagation environments to tackle the unfavorable channel conditions. 

\subsection{Principles of Reconfigurable Intelligent Surface}
RIS, as an artificial two-dimensional surface of electromagnetic (EM) materials, is designed to possess special properties that can transform the EM waves differently from the natural materials. 
In particular, the RIS is engineered to have periodic EM structures and be composed of a large number of passive sub-wavelength metallic scattering elements. 
With those specially designed scattering elements, the extreme values of effective permittivity and permeability coefficients can be electronically tuned to adjust the EM properties (e.g., phase shift) of incident waves. 
In addition, the controllable scattering elements can be implemented by variable lumped elements such as PIN diodes, varactor diodes, and micro-electro-mechanical systems (MEMS), or tunable materials such as ferro-electric thin film liquid crystal and graphene. 
Utilizing different materials for different spectrum results in different phase shift regions and energy loss.

The arbitrary re-direction of the incident signals is achieved by jointly controlling the phase shifts of all passive reflecting elements. 
Field-programmable gate array (FPGA) can be adopted as the micro-controller to simultaneously control the stimulus in each reflecting element such as switch state and the direct current bias voltage. Benefiting from the flexible control of individual phase shift, RIS is capable of achieving powerful functions, including perfect absorption, anomalous reflection, and wave manipulation. As a result, an RIS with real-time reconfigurability enables the software-controlled phase shift of incident signals, thereby making reconfigurable radio environments possible. 
  
\subsection{RIS-Empowered AirComp for Model Aggregation}
\begin{figure}[t]
  \centering
  \includegraphics[width=\columnwidth]{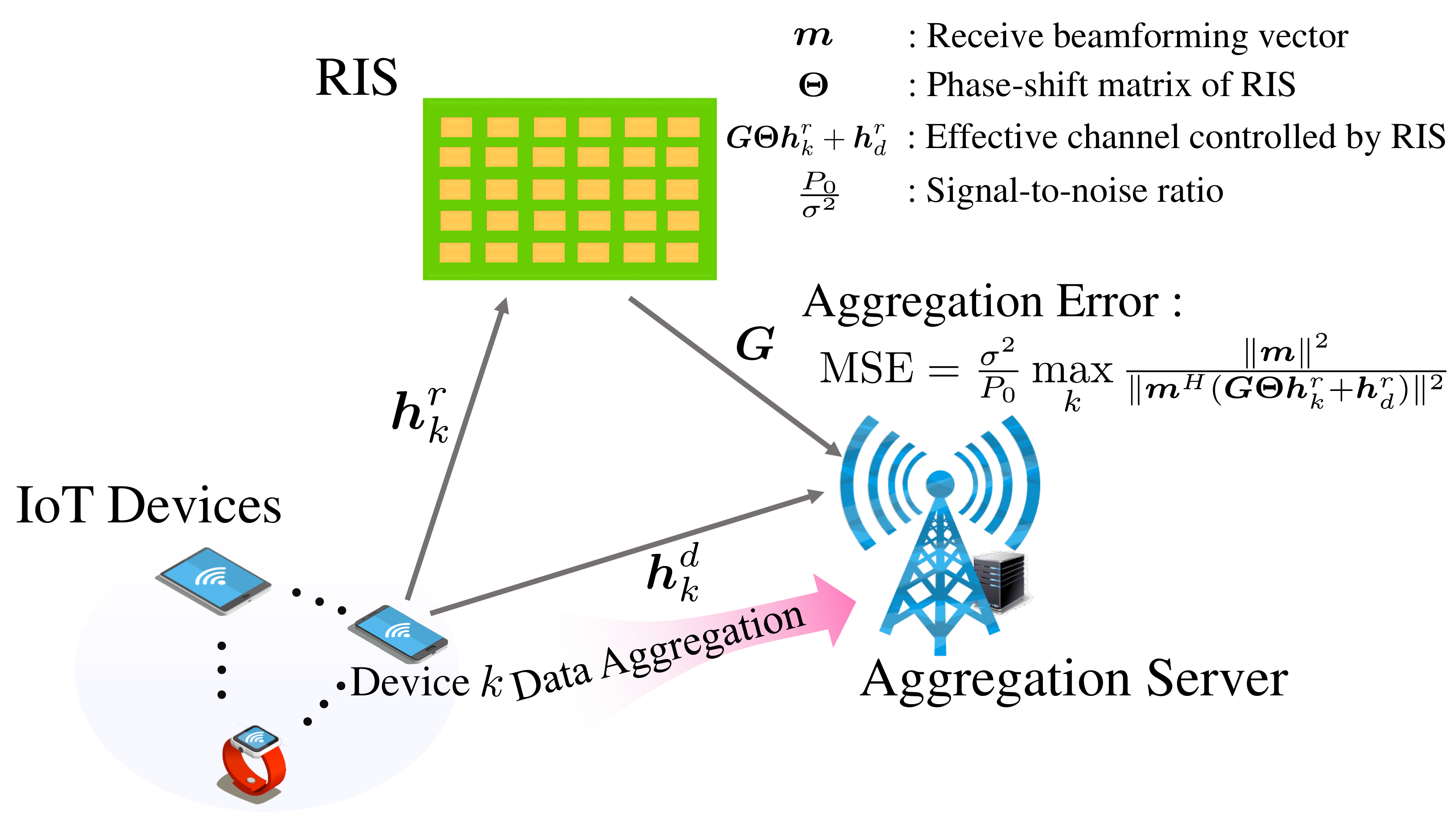}
  \caption{RIS-empowered  over-the-air computation for model aggregation.}
  \label{fig_RISsystem}
\end{figure}

For AirComp-based federated machine learning, minimizing the  model aggregation error that is quantified by MSE is crucial for enhancing the learning performance. 
It is worth noting that the MSE is critically dependent upon the channel conditions between the devices and the aggregation server. 
As the RIS is capable of achieving desired channel responses by enabling software-controlled phase shift, it is possible to reduce the MSE of model aggregation with the assistance of RIS, as illustrated in Fig. \ref{fig_RISsystem}. Compared with other smart radio environment approaches such as relay or active reflecting surfaces, RIS has the unique advantages of flexible deployment and low power consumption due to its passive nature.
Specifically, reducing the MSE of the aggregated global model improves the accuracy of the model prediction. 
Moreover, a smaller MSE makes it possible to select more devices at each communication round, which in turn speeds up the convergence of federated machine learning. 

With the great potentials, RIS-empowered model aggregation is highly advocated. 
To this end, the phase shifts of reflecting elements at the RIS should be optimized to minimize the MSE of model aggregation. 
The resulting MSE minimization problem requires the joint design of receive beamformer at the aggregation server and the phase shifts at the RIS. 
Such a problem is generally a computationally difficult nonconvex bi-quadratic programming problem \cite{jiang2019over}, i.e., nonconvex quadratic with respect to both the receive beamformer and RIS phase shift matrix. 
To decouple the optimization variables, the receive beamformer and the phase shifts are updated in an alternating fashion. 
In each alternation, it is required to solve a nonconvex quadratic constrained optimization problem, which can be addressed by developing a low-rank optimization approach with a DC algorithm \cite{jiang2019over}. 

\subsection{Illustrative Results}
\begin{figure}[t]
  \centering
  \includegraphics[width=\columnwidth]{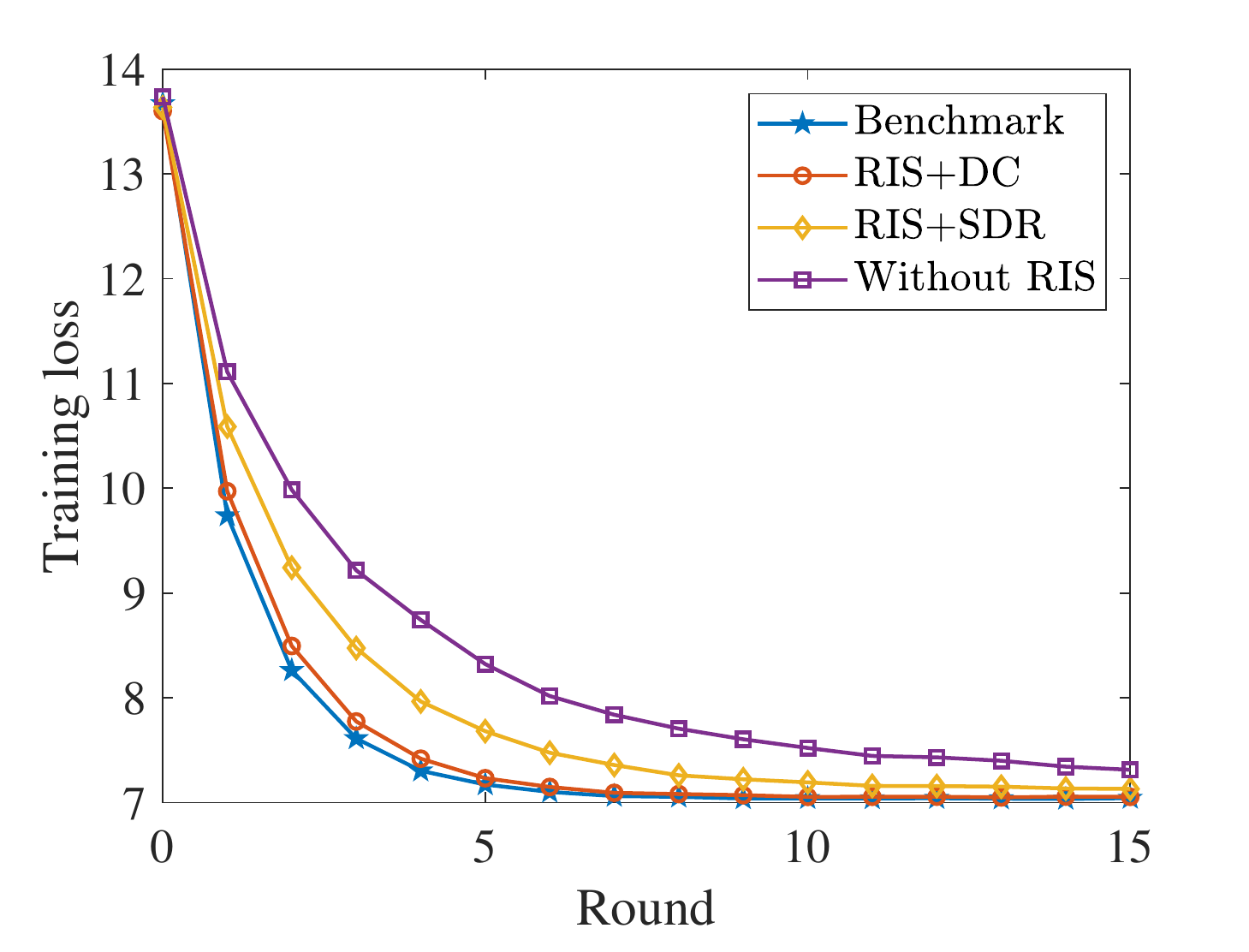}
  \caption{Training loss of RIS-empowered federated machine learning via AirComp. }
  \label{fig_loss}
\end{figure}

\begin{figure}[t]
  \centering
  \includegraphics[width=\columnwidth]{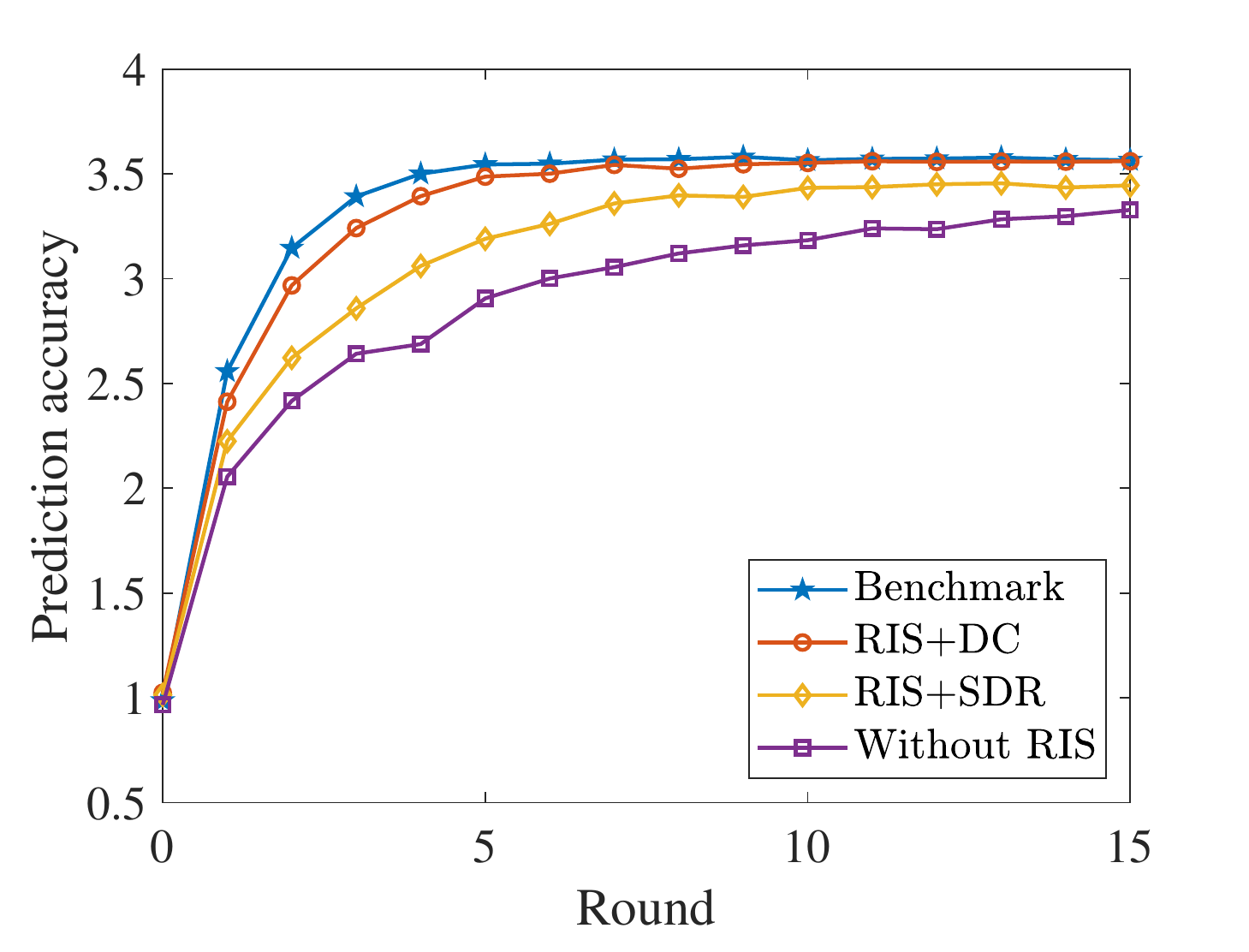}
  \caption{Prediction accuracy of RIS-empowered federated machine learning via AirComp. }
  \label{fig_prediction}
\end{figure}

We consider a federated machine learning system consisting of 20 single-antenna IoT devices to train a support vector machine (SVM) classifier with randomly split and deployed CIFAR-10 datasets. CIFAR-10, containing 10 different classes of objects, is a widely used image classification dataset. Suppose that all devices are involved in the model aggregation at each communication round. We evaluate the learning performance of the following three algorithms/settings in terms of training loss and prediction accuracy, i.e., the proposed DC approach for RIS-empowered AirComp model aggregation, the semidefinite relaxation (SDR) approach for RIS-empowered AirComp model aggregation, and the DC approach for AirComp model aggregation without RIS. 
We choose the benchmark as the learning performance of perfect model aggregation without any distortion. 
It is demonstrated in Figs. \ref{fig_loss} and \ref{fig_prediction} that 
RIS-empowered model aggregation achieves much lower training loss and higher prediction accuracy than that without RIS for AirComp-based federated machine learning thanks to the reduction of the model aggregation error. 

\section{Conclusions and Research Directions}
In this article, we discussed the potential of federated machine learning in addressing the critical challenges of intelligent IoT, so as to ensure data security, reduce network congestion, and make full use of distributed computation resources. 
We also identified two emerging technologies, i.e., AirComp and RIS, that can tackle the challenge of limited communication bandwidth during the model aggregation phase of federated machine learning. 

There are some interesting research directions for future work. The analog AirComp is generally difficult to be intergrated into the current digital communication systems. An interesting research direction is to study digital modulation schemes for AirComp to achieve fast model aggregation. In addition, quantization methods can be adopted to further reduce the communication bandwidth required by AirComp-based data aggregation. Another future work is to study discrete phase shifts of RIS for federated machine learning due to practical implementation concern. Also, \cite{huang2020reconfigurable} motivates us to develop deep reinforcement learning based approaches to achieve fast response to wireless radio environment by considerably reducing the computational complexity.
We hope this article will spur interests and further studies in the area of intelligent IoT with federated machine learning and RIS. 

\section{Acknowledgment}
This work was supported in part by the National Key Research and Development Program (Grant No. 2018YFA0701601) and the National Natural Science Foundation of China (NSFC) (Grant No. 61601290, 61971286 and 61771017).

\bibliographystyle{ieeetr}

\begin{IEEEbiography}{Kai Yang}
[S'16] (yangkai@shanghaitech.edu.cn) 
received the B.S. degree from the Dalian University of Technology. He is currently working toward the Ph.D. degree with the School of Information Science and Technology, ShanghaiTech University, Shanghai, China, also with the Shanghai Institute of Microsystem and Information Technology, Chinese Academy of Sciences, Shanghai, China, and also with the University of Chinese Academy of Sciences, Beijing, China.
   \end{IEEEbiography}\vskip 0pt plus -1fil
   \begin{IEEEbiography}{Yuanming Shi}
   [S'13, M'15] (shiym@shanghaitech.edu.cn)
received his B.S. degree from Tsinghua University and the Ph.D.
degree from The Hong Kong University of Science and Technology.
He is currently a tenured associate professor at the
School of Information Science and Technology, ShanghaiTech
University. 
   \end{IEEEbiography}\vskip 0pt plus -1fil
    \begin{IEEEbiography}{Yong Zhou}
   [S'13-M'16] (zhouyong@shanghaitech.edu.cn) received his Ph.D. Degree in
the Department of Electrical and Computer Engineering from University of
Waterloo in 2015. He is currently an Assistant Professor at the School of
Information Science and Technology, ShanghaiTech University, China. 
   \end{IEEEbiography}\vskip 0pt plus -1fil
   \begin{IEEEbiography}{Zhanpeng Yang}
    (zpyang@stu.xidian.edu.cn) will receive his B.S. degree from Xidian University.
He is currently a visiting student at the School of Information Science and
Technology, ShanghaiTech University.
    \end{IEEEbiography}\vskip 0pt plus -1fil
     \begin{IEEEbiography}{Liqun Fu}
  [S'08, M'11, SM'17] (liqun@xmu.edu.cn) received her Ph.D. Degree in Information
Engineering from The Chinese University of Hong Kong in 2010. She is currently
a Full Professor at the School of Informatics at Xiamen University. 
   \end{IEEEbiography}\vskip 0pt plus -1fil
   \begin{IEEEbiography}{Wei Chen}
    [S'05, M'07, SM'13] (wchen@tsinghua.edu.cn)
received his B.S. and Ph.D. degrees from Tsinghua University.
He was a visiting Ph.D. student at HKUST from 2005 to 2007.
He is currently a tenured professor with the Department of Electronic Engineering, Tsinghua University.
   \end{IEEEbiography}

\end{document}